\documentclass[10pt,twocolumn]{article}

\usepackage{graphicx}
\usepackage{dcolumn}
\usepackage{bm}
\usepackage{romannum}

\newcommand{\beq}{\begin{equation}}
	\newcommand{\eeq}{\end{equation}}
\newcommand{\bea}{\begin{eqnarray}}
	\newcommand{\eea}{\end{eqnarray}}

\title{Experimental proof of the reciprocal relation between spin Peltier and spin Seebeck effects in a bulk YIG/Pt bilayer}
\author{Alessandro Sola$^{(1)}$, Vittorio Basso$^{(1)}$, Michaela Kuepferling$^{(1)}$, Carsten Dubs$^{(2)}$, \\Massimo Pasquale$^{(1)}$
	\\
(1) Istituto Nazionale di Ricerca Metrologica, Strada delle Cacce 91, 10135, Torino, Italy\\
(2) INNOVENT e.V., Technologieentwicklung, Pr\"ussingstrasse. 27B, 07745 Jena, Germany
}

\begin{document}

\maketitle

\begin{abstract}
We verify for the first time the reciprocal relation between the spin Peltier and spin Seebeck effects in a bulk YIG/Pt bilayer. Both experiments are performed on the same YIG/Pt device by a setup able to accurately determine heat currents and to separate the spin Peltier heat from the Joule heat background. The sample-specific value for the characteristics of both effects measured on the present YIG/Pt bilayer is $(6.2 \pm 0.4)\times 10^{-3} \,\, \mbox{KA$^{-1}$}$. In the paper we also discuss the relation of both effects with the intrinsic and extrinsic parameters of YIG and Pt and we envisage possible strategies to optimize spin Peltier refrigeration.
\end{abstract}

\section{Introduction}
\label{Sec:Int}

The reciprocal relations of thermodynamics are a fundamental tool to analyze and understand the physics of transport phenomena \cite{Callen-1985}. Since the beginning of the 19th century it was clear to Jean Charles Athanase Peltier and later demonstrated by Lord Kelvin, that for a material at a given absolute temperature $T$ a relation exists between the Seebeck coefficient $\epsilon$ (given by the ratio between the measured electromotive force and the applied temperature difference) and the Peltier coefficient $\Pi$ (the ratio between the measured heat current and the applied electric current ): $\Pi = \epsilon T$ \cite{Miller-1960, Goldsmid-2001}. This remarkable reciprocity was later found to be part of a wider set of relations, as theoretically demonstrated by Onsager under the assumption of the reversibility of the microscopic physical processes governing macroscopic non-equilibrium thermodynamic effects \cite{Onsager-1931}.

Reciprocal relations can also be used to analyze transport phenomena which involve not only the electric charge and the heat, but also the spin. Spincaloritronic phenomena \cite{Bauer-2012, Boona-2014,Yu-2017} can provide additional tools to the field of spintronics, envisioned to be a faster and lower energy consuming alternative to classical electronics \cite{Ansermet-2010}. One of the key building blocks for "spintronic circuits" is the spin battery, a device which can drive a spin current into an external circuit. Spin batteries are fundamental for spintronic devices and may be developed exploiting spincaloritronic effects \cite{Yu-2017b}. Spincaloritronic devices may also be used in the development of novel  thermoelectric heaters/coolers operating at the microscale \cite{Gonzalez-2013,Uchida-2016}. 
The idea of a reciprocity between heat and spin was initially proposed and proven for metals where the spin current is carried by electrons \cite{Dejene-2014}, but such a reciprocal relation cannot be easily proven in the case of ferrimagnetic insulators where the spin current is carried by thermally excited spin waves \cite{Flipse-2014}. A typical device, where spincaloritronic effects are found and can be exploited for experiments, is a bilayer made using a ferrimagnetic insulator (e.g. yttrium iron garnet, YIG) and a non magnetic metal with a strong spin-orbit coupling (e.g. platinum, Pt) \cite{Uchida-2010c}. In these devices the spin current is generated longitudinally (along the $x$ axis), normal to the film surface.

In the case of devices which exhibit the spin Peltier effect (SPE) a longitudinal ($x$ axis) heat current is generated, caused by the flow of a transverse ($y$ axis) electric current in the Pt layer \cite{Flipse-2014}. Conversely, in the case of the spin Seebeck effect (SSE) a transverse ($y$ axis) electric voltage is generated in the Pt layer and caused by the longitudinal temperature gradient parallel to the spin current ($x$ axis) \cite{Uchida-2010}. 
  
Although experimental evidence of both effects has been already obtained \cite{Uchida-2014b, Daimon-2017}, the quantitative demonstration of their reciprocity, the relation between the two effects and the connection with intrinsic properties of the layers, has yet to be proven \cite{Flipse-2014, Daimon-2016, Uchida-2017, Itoh-2017}.

To this end here we provide the first experimental evidence of the reciprocal relation between the thermal and the electric quantities associated to the SPE and the SSE in a bulk YIG/Pt bilayer. The relation for a YIG/Pt bilayer has the following form \cite{Basso-2016, Basso-2018}

\beq
\frac{- \Delta T_{SP,x}}{I_{e,y}}  = \frac{\Delta V_{e,y}}{I_{q,x}} T
\label{EQ:reciprocity1}
\eeq

\noindent The left hand side of Eq.(\ref{EQ:reciprocity1}) refers to the SPE: $\Delta T_{SP,x}$ is the temperature difference generated across the YIG layer as consequence of the electric current $I_{e,y}$ flowing in the Pt film. The right hand side of Eq.(\ref{EQ:reciprocity1}) refers to the SSE: $\Delta V_{e,y}$ is the voltage drop across the Pt film caused by the heat current $I_{q,x}$ flowing across the device and $T$ is the average temperature of the YIG. The experiments are performed measuring the SPE and the SSE on the same YIG/Pt device. The experimental value which represents, within the uncertainty, both the SPE and the SSE response of the specific device is $(6.2 \pm 0.4)\times 10^{-3} \,\, \mbox{KA$^{-1}$}$.

\begin{figure*}[ht]
\centering
\includegraphics[width=16.5cm]{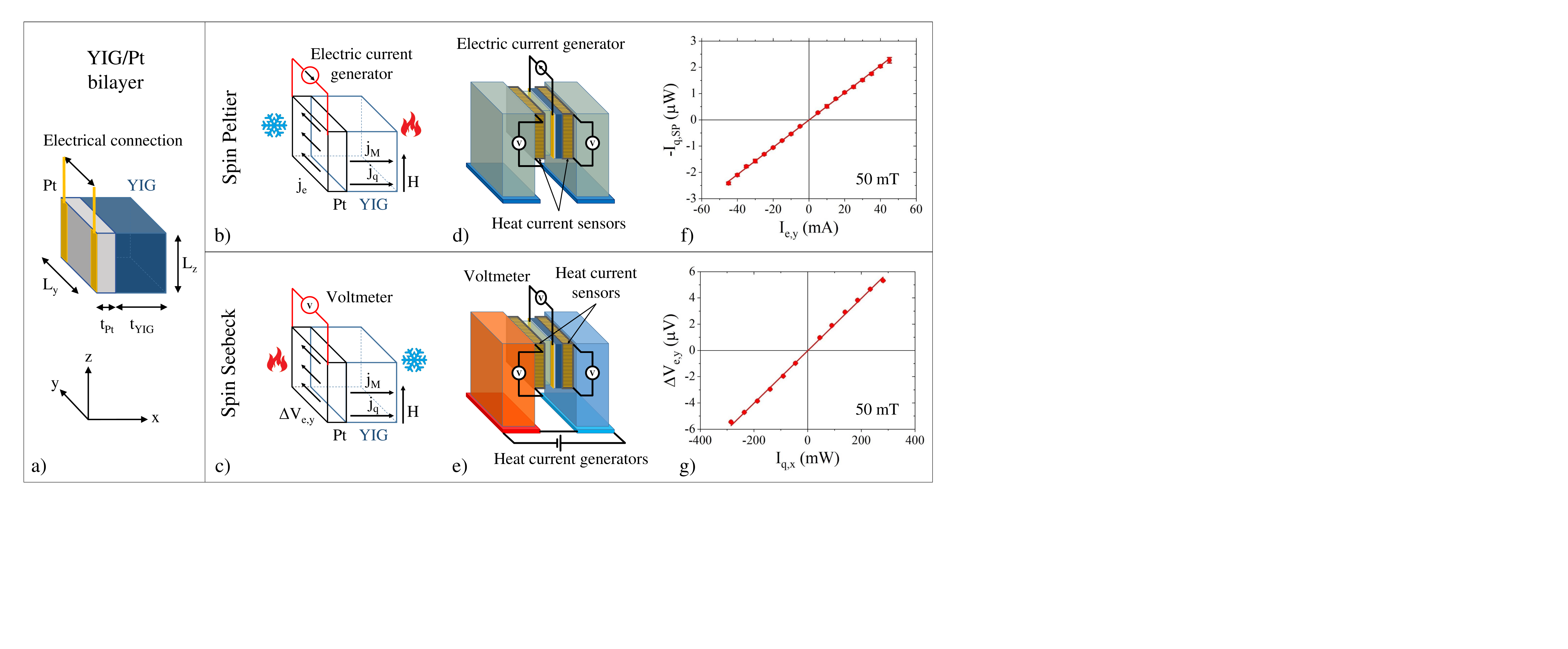}
\caption{a) Geometry of the YIG/Pt bilayer with $t_{YIG}=0.545$ mm, $L_y=4.95$ mm, $L_z=3.91$ mm and $t_{Pt}=$ 5 nm. b) and c) schemes of the device in the SPE and SSE configurations. Heat currents and magnetic moment currents are along $x$ (longitudinal direction), electric effects are along $y$ (transverse direction), the magnetic field and the magnetization are along $z$. d) and e) sketches of the experimental setup for SPE and SSE, respectively. f) Experimental result of the SPE heat current, $-I_{q,SP}$, as function of the electric current $I_{e,y}$ at positive magnetic saturation. g) Experimental result of the SSE voltage $\Delta V_{e,y}$ as function of the heat current $I_{q,x}$ at positive magnetic saturation.}
\label{FIG:Fig1}
\end{figure*}

\section{Results}
\label{Sec:Res}

\subsection{Device geometry and measurement principle}
\label{ssec:meas}

The geometry of the bulk YIG/Pt bilayer is shown in Fig.\ref{FIG:Fig1}a. The device is composed of a bulk YIG parallelepiped with a thin film of Pt sputter deposited on one side. The temperature gradient $\nabla_x T$ and the heat current density $j_{q,x}$ are directed along the $x$ axis. The electric voltage $\Delta V_{e,y}$ and the electric current density $j_{e,y}$ are directed along the $y$ axis. The magnetic moment current density, $j_M$, is along the $x$ direction and transports magnetic moments directed along the $z$ direction. The magnetic field $H$ and the magnetization $M$ of the bulk-YIG are also directed along the $z$ axis (Fig.\ref{FIG:Fig1}b and c).


The measurement setup for both SPE and SSE is shown in the sketches of Figs.\ref{FIG:Fig1}c and e. The YIG/Pt device is sandwiched between two thermal reservoirs (held at $T_h$ and $T_c$ respectively) in order to form a closed thermal circuit.  The temperature of the two reservoirs can be externally controlled and the two heat currents between the device and each of the reservoirs are measured simultaneously by sensitive heat flux detectors. The heat flux technique is chosen to avoid measurement uncertainties due to the hardly reproducible realization of thermal contacts  \cite{Sola-2015, Sola-2017, Bougiatioti-2017, Prakash-2018}. In order to minimize heat leakages in the thermal circuit, the whole setup is operated in vacuum. Technical, constructional and measurement details are reported in the Methods section.

\subsection{Spin Peltier effect}
\label{ssec:meas}
%

\begin{figure*}[ht]
	\centering
	\includegraphics[width=8cm]{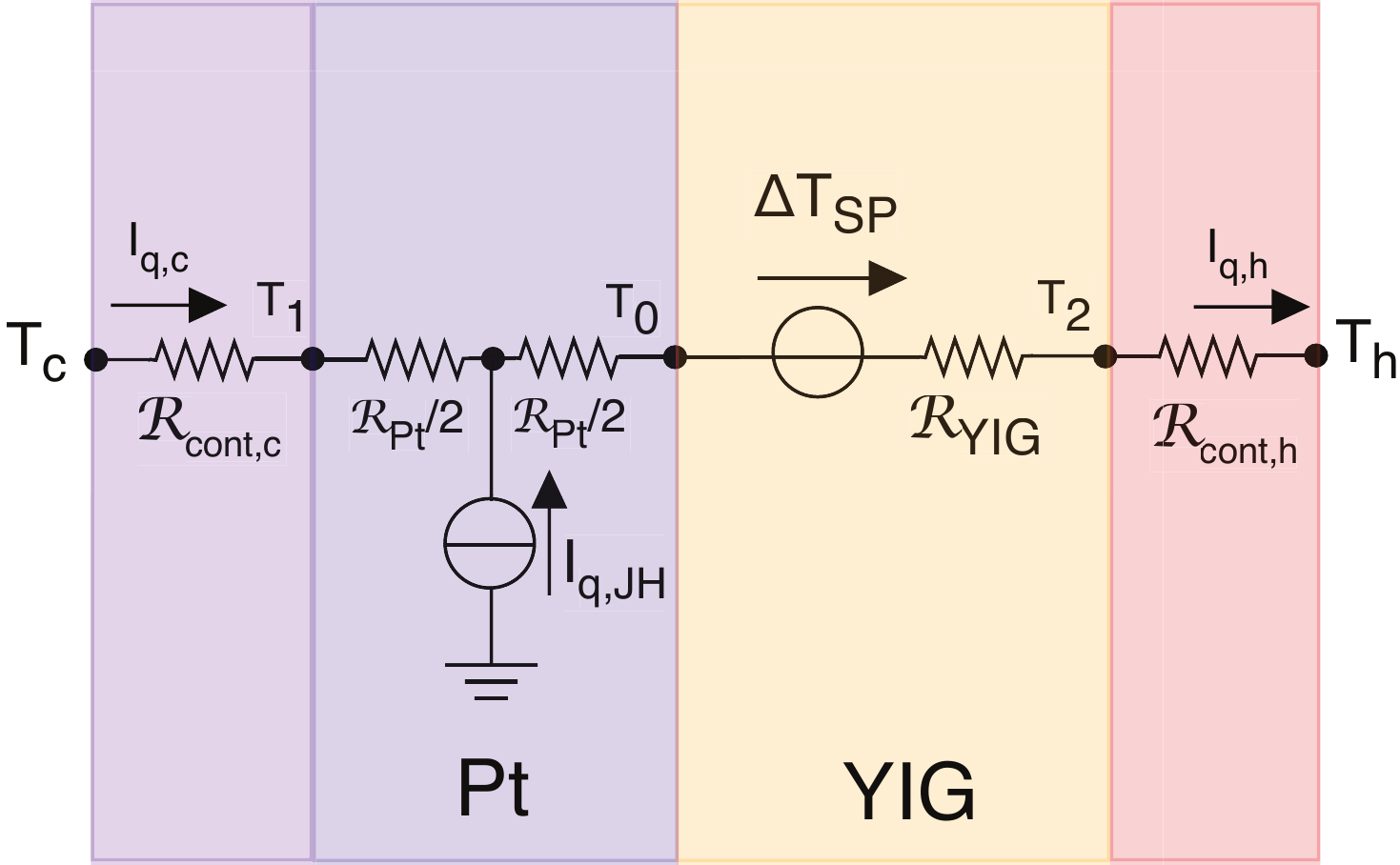}
	\caption{Equivalent thermal circuit of the SPE measurements of the YIG/Pt bilayer. The YIG layer is represented by the SPE generator, $\Delta T_{SP}$, and by the thermal resistance $\mathcal{R}_{YIG}$. The Pt layer has a Joule heat current source, $I_{q,JH}$, and a thermal resistance $\mathcal{R}_{Pt}$. The circuit includes two thermal contact resistances $\mathcal{R}_{cont,c}$ and $\mathcal{R}_{cont,h}$ taking into account both the thermal resistance of the contacts and the presence of the heat flux sensors. The difference $I_{q,h} - I_{q,c} = I_{q,JH}$ provides the Joule heat. In isothermal conditions, with $T_h=T_c=T$, we have $\Delta T_{SP} = \mathcal{R} I_{q,s} +I_{q,JH}(\mathcal{R}_h-\mathcal{R}_c)/2$ where $\mathcal{R}_{h} =  \mathcal{R}_{Pt}/2 + \mathcal{R}_{YIG} + \mathcal{R}_{cont,h}$, $\mathcal{R}_{c} = \mathcal{R}_{cont,c} + \mathcal{R}_{Pt}/2$, $I_{q,s} = (I_{q,h} + I_{q,c})/2$ is the half sum and $\mathcal{R} = \mathcal{R}_h+\mathcal{R}_c$ is the total resistance (see Supplementary material).}
	\label{FIG:Eq}
\end{figure*}

In the SPE, an electric current $I_{e,y}$ flowing in the Pt layer generates a magnetic moment current $j_M$ along the $x$ direction as a result of the spin Hall effect \cite{Sinova-2015}. The adjacent ferrimagnetic bulk YIG here acts as a passive component and shows a longitudinal ($x$ axis) heat current associated to the magnetic moment current injection \cite{Rezende-2016, Basso-2016b, Nakata-2017}. The measured heat current is however also including the Joule heat contribution generated by the electric current flowing in the Pt layer. In order to separate the Joule and spin Peltier contributions, their intrinsic differences have to be exploited: the spin Peltier signal increases linearly with the $I_{e,y}$ current and changes sign when the magnetization (along the $z$ axis) or the $I_{e,y}$ are inverted (odd parity), while Joule heating is proportional to $I^2_{e,y}$ and does not change sign under an inversion (even parity) \cite{Flipse-2014,Daimon-2016}. 

\begin{figure*}[ht]
	\centering
	\includegraphics[width=16.5cm]{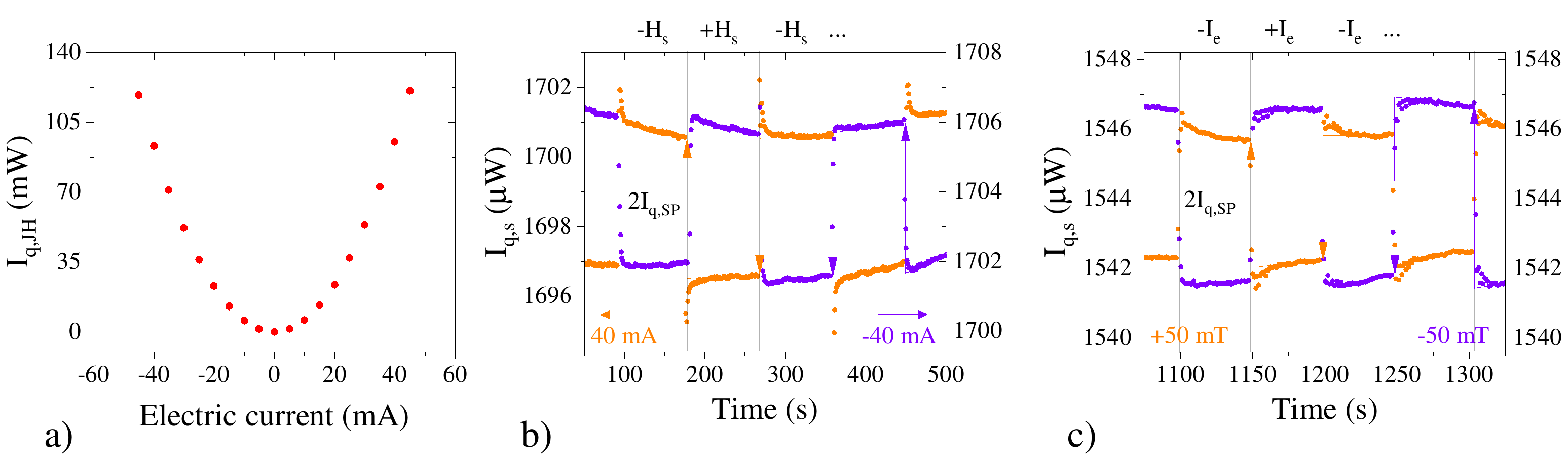}
	\caption{Heat current signals measured on the YIG/Pt bilayer during the SPE experiment. a) Joule heat signal given by the difference $I_{q,JH} = I_{q,h} - I_{q,c}$. b) SPE by magnetic field inversion. Half sum signal $I_{q,s} = (I_{q,h} + I_{q,c})/2$ caused by a rectangular waveform of the magnetic field $|\mu_0H_{s}|=50$ mT, for two steady values of electric current ($\pm$ 40 mA, orange/purple). c) SPE by electric current inversion. $I_{q,s}$ caused by a rectangular waveform of the  electric current $|I_{e,y}|=40$ mA, for two steady values of magnetic field ($\pm$ 50 mT, orange/purple). In both cases the spin Peltier signal $I_{q,SP}$ is obtained as half of the variation at the inversion $I_{q,SP} = \Delta I_{q,s}/2$. The variation $\Delta I_{q,s}$ is taken at the inversion instant and is computed from the extrapolation of the linear fit taken a few seconds after the inversion. This method permits to avoid the contributions of spurious induced voltage spikes after the inversions. The values reported in Fig.\ref{FIG:Fig1}f) are the result of the average of 10 inversions.}
	\label{FIG:Fig3}
\end{figure*}

The thermal problem of the SPE can be represented by the equivalent circuit of Fig.\ref{FIG:Eq} \cite{Basso-2018} (see also Supplementary material). In adiabatic conditions the SPE corresponds to the direct measurement of $\Delta T_{SP}$, the temperature difference generated between the two faces of the bulk YIG sample. Previous experimental work has succeeded to extract $\Delta T_{SP}$ out of the Joule heat component by using an AC technique \cite{Flipse-2014,Daimon-2016}. Here, to test the reciprocal relation in a stationary state, we employ a DC technique in which we set isothermal conditions at the thermal baths, $T_h = T_c = T$, and measure simultaneously the two heat currents: $I_{q,c}$ and $I_{q,h}$. The difference of the heat flux signals $I_{q,h} - I_{q,c} = I_{q,JH}$ provides the Joule heat only (see Fig.\ref{FIG:Fig3} a), while the half sum $I_{q,s}=(I_{q,h} + I_{q,c})/2$ contains the SPE signal (see Fig.\ref{FIG:Fig3} b and c).

We first detect the SPE generated by setting a constant electric current (i.e. $I_{e,y}$= 40 mA, see Fig.\ref{FIG:Fig3}b, orange points) and periodically inverting the magnetic field. The half sum signal $I_{q,s}$ has a change of $\pm2I_{q,SP}$ at each inversion. Equivalently, when the sign of the electric current in the Pt film is changed (i.e. $I_{e,y}$= - 40 mA, Fig.\ref{FIG:Fig3} b, purple points), a sign inversion of the change $\mp 2I_{q,SP}$ occurs. The field inversion allows to detect the small contribution of the spin Peltier heat current (a few $\mu$W) superimposed to the Joule heating background (a few mW).

As a second step we measure the SPE signal $I_{q,s}$ at the constant magnetic field $\mu_0H_s$ = +50 mT while the applied current $I_{e,y}$=$\pm$40 mA is periodically inverted, Fig.\ref{FIG:Fig3}c, orange points. Conversely when we apply $\mu_0H_s$ = -50 mT and the applied current $I_{e,y}$=$\pm$40 mA is inverted, we obtain the curve of Fig.\ref{FIG:Fig3}c, purple points. The current inversion method provides the same results, within the uncertainty, as the magnetic field inversion one, provided one takes into account the presence of small spurious offset signals as discussed in the Methods section. This second method allows to detect the SPE signal as function of the applied magnetic field. Therefore it permits the determination of the hysteresis loop of YIG \cite{Uchida-2015} (more details about this experiment are reported in Supplementary materials.).

At the saturating magnetic field $\mu_0H_s$ = +50 mT, by applying different values of $I_{e,y}$ and by deriving the corresponding values of $-I_{q,SP}$ as shown in Fig.\ref{FIG:Fig3}b and c, we are able to obtain the linear relation between the SPE heat current and the electric current data of Fig.\ref{FIG:Fig1}f . By a linear fit we find

\beq
\frac{-I_{q,SP}}{I_{e,y}} =(5.1 \pm 0.3)\times 10^{-5} \,\,  \mbox{WA$^{-1}$}
\eeq

\noindent The thermal resistance $\mathcal{R}$ of the whole stack consisting of sensors, sample and additional thermal contacts (e.g. thermal paste or thermally conducting layers) is experimentally measured by setting a heat current value and measuring the temperatures of the two thermal reservoirs by two thermocouples. The result is $\mathcal{R} = (119 \pm 2 )$ KW$^{-1}$. With $\Delta T_{SP} = \mathcal{R} I_{q,SP}$, the measured spin Peltier coefficient is

\beq
\frac{-\Delta T_{SP}}{I_{e,y}} =(6.1 \pm 0.4)\times 10^{-3} \,\, \mbox{KA$^{-1}$}
\label{EQ:DTSPE}
\eeq

\noindent and since the current density flowing in the Pt film is $j_{e,y} = I_{e,y}/(t_{Pt}L_{z})$ with $L_z = 3.9$ mm and $t_{Pt}=$5 nm we are finally able to obtain the intrinsic SPE coefficient:

\beq
\frac{-\Delta T_{SP}}{j_{e,y}} =(1.19 \pm 0.08)\times 10^{-13} \,\, \mbox{Km$^2$A$^{-1}$}.
\eeq

\subsection{Spin Seebeck effect}
\label{ssec:meas}

In the SSE experiment the bulk YIG is the active layer which generates a magnetic moment current when subjected to a temperature gradient, while the Pt is the passive layer in which the injected magnetic moment current is converted into a transverse electric potential. The SSE signal measured across the Pt film $\Delta V_{e,y}$ changes sign when the  magnetization of the YIG is inverted in sign (odd parity). 

The measurements of the SSE is performed by setting a value of the heat current $I_{q,x}$ traversing the YIG/Pt device and measuring the consequent voltage $\Delta V_{e,y}$ found on the Pt layer when the YIG layer is at magnetic saturation. The set of values obtained  $\Delta V_{e,y}$ versus the heat current $I_{q,x}$ is shown in Fig.\ref{FIG:Fig1}f and a linear fit gives 

\beq
\frac{\Delta V_e}{I_{q,x}} = (2.1 \pm 0.1) \times 10^{-5} \,\, \mbox{VW$^{-1}$}
\eeq

\noindent A geometry independent (intrinsic) spin Seebeck coefficient can be defined as $\nabla_y V_e/j_{q,x}$ with $\nabla_y V_e = \Delta V_e/L_{e,y}$ and $ j_{q,x} = I_{q,x}/A_q$ with $A_{q} = L_{e,y} \times L_{z}$. With $L_{e,y} = 4.17$ mm, the dimension of the Pt electrode used to detect the voltage drop, and $L_{z}=3.91$ mm we have

\beq
\frac{\nabla_y V_e}{j_{q,x}}= (8.2 \pm 0.3) \times 10^{-8} \,\, \mbox{VmW$^{-1}$}
\eeq

\noindent Finally the spin Seebeck coefficient $S_{SSE} = \nabla_y V_{e}/\nabla_x T$, given by the ratio between the transverse gradient of the electric potential $\nabla_y V_{e}$ in Pt and the longitudinal gradient of the temperature $\nabla_x T$ in YIG (as it is often defined in literature) can be obtained

\beq
S_{SSE} = - \frac{\nabla_y V_e}{j_{q,x}} \kappa_{YIG} = - (5.4 \pm 0.2) \times 10^{-7} \,\, \mbox{VK$^{-1}$}
\eeq

\noindent using the bulk value of the thermal conductivity of YIG: $\kappa_{YIG} = 6.63$ Wm$^{-1}$K$^{-1}$ \cite{Hofmeister-2006}. This result concurs with another experimental bulk YIG $|S_{SSE}| \simeq 4 \times 10^{-7}$ VK$^{-1}$  \cite{Uchida-2014}.

\section{Discussion}
\label{sec:res}

The microscopic and physical origin of the spin Peltier and of the spin Seebeck effects have been investigated in detail \cite{Chotorlishvili-2013, Cunha-2013, Etesami-2014, Rezende-2014, Ritzmann-2015, Cornelissen-2016, Rezende-2016, Geprags-2016, Nakata-2015b, Basso-2016b, Nakata-2017, Ohnuma-2017, Saslow-2017, Prakash-2018}. These two effects are the results of two independent physical mechanisms. In YIG the presence of a spin current or, more generally, of a magnetic moment current, carried by thermally excited spin waves, is accompanied by a heat current. In Pt the longitudinal spin polarized current is associated with a transverse electric effect due to the inverse spin Hall effect, described by the spin Hall angle $\theta_{SH}$.
At the interface between YIG and Pt the spin current is partially injected from one layer into the other  \cite{Saitoh-2006, Bender-2015}. By adopting the thermodynamic description of Johnson and Silsbee \cite{Johnson-1987} further developed in Refs.\cite{Rezende-2016, Basso-2016, Basso-2016b, Nakata-2017, Basso-2018}, the thermomagnetic effects in YIG are described by means of the thermomagnetic power coefficient $\epsilon_{YIG}$, that has an analogous role to the thermoelectric power coefficient $\epsilon$ of thermoelectrics.
At the interface, the passage of the magnetic moment current, due to diffusion, is mainly determined by the magnetic moment conductances per unit surface area, $v_M$, of the two layers. Basing on these ideas, it has been possible to work out the reciprocal relation relating the spin Seebeck and spin Peltier effects with the intrinsic and extrinsic parameters of the bilayer. The expression is \cite{Basso-2016, Basso-2018}

\beq
\frac{-\Delta T_{SP}}{j_e} \frac{1}{T}  = \frac{\nabla_{y} V_{e}}{j_{q,x}} t_{Pt} = \theta_{SH} \mu_0 \left(\frac{\mu_B}{e} \right) \frac{1}{v_p} \frac{ \epsilon_{YIG} \sigma_{YIG}}{\kappa_{YIG}}
\label{EQ:reciprocity2}
\eeq

\noindent Eq.(\ref{EQ:reciprocity2}) contains two equal signs. The equal sign at the left is between the SPE and SSE measured quantities and we find the temperature difference between the two faces of YIG, $\Delta T_{SP}$, of the SPE generated by the electric current density $j_{e,y}$ and the transverse gradient of the electric potential in Pt, $\nabla_y V_{e}$, of the SSE, generated by the heat current density, $j_{q,x}$, in YIG. The equal sign at the right refers to the relation of both SPE and SSE to intrinsic coefficients. In addition to the expected parameters: $\theta_{SH}$, the spin Hall angle of Pt and $\epsilon_{YIG}$, the thermomagnetic power coefficient of YIG, we have: $\sigma_{YIG}$, the magnetic moment conductivity of YIG, and $v_p$, the magnetic moment conductance per unit surface area of the YIG/Pt interface and $\kappa_{YIG}$, the thermal conductivity of the YIG. $\mu_0$ is the magnetic constant, $\mu_B$ is the Bohr magneton and $e$ is the elementary charge. $v_p$ depends on the intrinsic conductances,  $v_M$ of both YIG and Pt and on the ratio $t/l_M$ between the thickness $t$ and the magnetic moment diffusion length $l_M$, for each layer. The expression of $v_p$ is derived in Ref.\cite{Basso-2018} and reported in the Supplementary material.

By just taking the first equal sign of Eq.(\ref{EQ:reciprocity2}) and integrating over the size of the bilayer device we have

\beq
\frac{-\Delta T_{SP}}{I_{e,y}/(L_{z} \cdot t_{Pt})} \frac{1}{T} = \frac{\Delta V_{e,y}/L_{q,y}}{I_{q,x}/(L_{q,y} \cdot L_{z})} t_{Pt}
\label{EQ:reciprocity_densities}
\eeq

\noindent At the left hand side $(L_{z} \cdot t_{Pt})$ is the area where the electric current flows. At the right hand side $(L_{q,y} \cdot L_{z})$ is the area of the thermal contact and corresponds to the region where the spin Seebeck effect rises.  The previous relation can be simplified, leading to Eq.(\ref{EQ:reciprocity1}) and permitting a direct test of the reciprocity by using the experimental values. By taking the average temperature $T = (298\pm 2)$ K, the spin Seebeck experiment gives

\beq
\frac{\Delta V_{e,y}}{I_{q,x}} T = (6.3 \pm 0.3)\times 10^{-3} \,\, \mbox{KA$^{-1}$}
\label{EQ:reciprocitySSE}
\eeq

\noindent which is in excellent agreement with the spin Peltier value of Eq.(\ref{EQ:DTSPE}) and verifies experimentally the reciprocal relation, Eq.(\ref{EQ:reciprocity1}), between spin Seebeck and spin Peltier effects.

Once the reciprocity is verified, we can take the second equal sign of Eq.(\ref{EQ:reciprocity2}) and obtain an experimental value that can be compared with the theoretical coefficients. By labeling the value taken from equation (\ref{EQ:reciprocity2}) as $K_{YIG/Pt}$ we find from the experiments

\beq
K_{YIG/Pt} = (4.0 \pm 0.3)\times 10^{-16} \,\, \mbox{m$^2$A$^{-1}$}
\eeq

\noindent We now compare the obtained value with the coefficients known from the literature. By using $\mu_{0}\sigma_{YIG} = v_{YIG}l_{YIG}$ and employing the values of the coefficients determined in previous works \cite{Basso-2016} ($\epsilon_{YIG} \simeq -10^{-2}$ TK$^{-1}$ and $\theta_{SH}=-0.1$) we can quantify the only missing parameter, $v_p/v_{YIG}$, as $v_p/v_{YIG}\simeq 9 $. This value is compatible with the transmission of the magnetic moment current between the two layers as determined by the intrinsic diffusion lengths $l_M$, by the thicknesses $t$ and by the intrinsic conductances $v_M$ of YIG and Pt. By using the expression for $v_p$ \cite{Basso-2018} with $l_{Pt} \simeq 7.3$ nm and $l_{YIG} \sim 0.4 \, \mu$m and with $v_{YIG} \sim v_{Pt}$ we obtain $v_p /v_{YIG}= 8$ which is reasonably close to the measured value.

We are therefore able to evaluate the cooling potentiality of the spin Peltier effect. If we consider the YIG/Pt device able to operate in adiabatic conditions ($I_{q,c}=0$), the temperature change across the device $\Delta T$ will be

\beq
\Delta T = \Delta T_{SP} - \mathcal{R}_{YIG} I_{q,JH}
\eeq

\noindent where we have assumed $\mathcal{R}_{Pt} \ll \mathcal{R}_{YIG}$ and $\mathcal{R} \simeq \mathcal{R}_{YIG}$. By taking the specific device studied in this paper, the electric current which is maximizing $\Delta T$ is $I_{e,y} = - 12.4\, \mu$A giving a maximum temperature change of $\Delta T = 3.8 \times 10^{-8}$ K. This value appears so small to discourage any attempt to employ the SPE in practice. However the verification of the validity of the thermodynamic theory for the SPE and SSE (Eq.(\ref{EQ:reciprocity2})) offers, in future perspective, the possibility to design and optimize spin Peltier coolers and spin Seebeck generators going beyond the specific bilayer sample used in this experimental study. Work is in progress along this line, however two main preliminary comments are already possible at this stage. The first is to identify the YIG and Pt thicknesses that would optimize the effects. The answer comes from the fact that both materials are active over a thickness of the order of the diffusion length $l_M$. Therefore promising devices would have $t_{YIG} \sim l_{YIG}$ and $t_{Pt}\sim l_{Pt}$. The second is that, as for thermoelectrics, the thermomagnetic YIG is characterized by a figure of merit $\zeta_T = \epsilon_{YIG}^2\sigma_{YIG} T /\kappa_{YIG}$, which is indeed very small at room temperature, $\zeta_T\simeq 4 \times 10^{-3}$ \cite{Basso-2018}. However it is expected that the $\zeta_T$ parameter could improve in the temperature range between 50 and 100 K where experiments \cite{Kikkawa-2015} have reported a much larger SSE (almost a factor 5) than the room temperature value. Finally it is worth to mention that both improvements could further benefit by cascading several devices in thermal series \cite{Ramos-2015}. For example, with $\zeta_T \sim 1$ and using a multilayer with an appropriate compensation of the Joule heat of each layer by using variable cross sections, one could obtain up to an effective $\Delta T = 20$ K for a device containing $\sim 10^3$ junctions. Work along this line has already progressed and future improvements are expected \cite{Uchida-2017}.

In summary, we investigated experimentally both the SPE and the SSE in a bulk YIG/Pt device. The thermal observables of these experiments are investigated by means of heat current measurements. This introduces a novel technique for the SPE characterization of a given sample in the DC regime. The experimental results of the SPE and SSE are used to verify for the first time the reciprocal relation between the two. The relation between both effects with the intrinsic and extrinsic parameters of YIG and Pt bilayer offers the possibilities for a more in-depth investigation of the applicability of spincaloritronic devices.

\appendix

\section{Methods}
\label{ssec:set}

\subsection{Device and experimental setup}

The YIG/Pt device employed is made of a bulk yttrium iron garnet (YIG) single crystal prepared by crystal growth in high-temperature solutions applying the slow cooling method \cite{Gornert-1984}.
Single crystals which nucleate spontaneously at the crucible bottom and grow to several centimetre sizes have been separated from the solution by pouring out the residual liquid. Afterwards, one crystal was cut parallel to one of its facet’s to prepare slices and parallelepipeds of the following dimensions: $L_y=4.95$ mm, $L_z=3.91$ mm, $t_{YIG}=0.545$ mm. Carefully grinding and polishing result in a sample with optical smooth surfaces ($R_{q}=0.4$ nm obtained by AFM). After polishing both sides, a thin film of Pt ($t_{Pt}$ about 5 nm) was sputtered on the top of one of the $L_y \times L_z$ surfaces at University of Loughborough (United Kingdom). Subsequently two 100 nm thick gold electrode strips for electric contacts were deposited. The inner distance between the electrodes is equal to $L_{e,y}=4.17$ mm. The Au contacts on the Pt layer are electrically connected to 40 $\mu$m diameter platinum leads by silver paste. The SSE voltage is measured by means of a Keithley 2182 nanovoltmeter while the SPE electric current is generated by a Keithley 2601 source meter. 

In order to avoid heat leakages all measurements are performed under vacuum (1.6$\times10^{-4}$ mbar) by means of a turbomolecular pump. The heat current sensors are miniaturized Peltier cells (5 mm $\times$ 5 mm and 1.9 mm of thickness, RMT Ltd model 1MD04-031-08TEG), calibrated according to the procedure described in Ref.\cite{Sola-2017}. The characteristic $I_q = - S_p V_p$ of both heat sensors are described by $S_p = 0.97 \pm 0.01$ V/W, where $I_q$ is the heat current traversing the sensor and $V_p$ is the voltage measured by a nanovoltmeter. The thermal reservoirs are two brass parallelepipeds (1 cm $\times$ 1 cm $\times$ 2 cm) to which one face of each heat sensor is glued with silver paste. The other face of each sensor is clamping the YIG/Pt device. We use an aluminum nitride slab (3 mm $\times$ 3 mm $\times$ 4 mm) with a large nominal thermal conductivity (140 - 180 Wm$^{-1}$K$^{-1}$) as geometrical adapter to thermally connect the Pt side of the device with the corresponding heat sensor. Thin layers of silicon based thermal grease are used to ensure uniform thermal conductivity through the sections.

\subsection{Offset subtractions during SPE and SSE measurements}

In the case of the electric current inversion method, the spin Peltier heat signals is affected by a spurious heat current offset, $I_{q,off}$, (a few percent of the total) that is not present in the magnetic field inversion method and should be therefore subtracted. The reason for this spurious heat current offset is that a conventional Peltier effect arises in presence of different metal contacts in the measurement setup (i.e. the contact between Pt and Au layers and the contacts with the electrical leads). These contacts would give a transverse (along $y$) heat flux, however the presence of an even very small transverse heat leakage may contribute to a small longitudinal contribution.
This spurious conventional Peltier effect $I_{q,off}$ presents, with respect to the current inversion, the same odd parity as the spin Peltier signal. 
When using the odd parity of the SPE under magnetic field inversion the spurious effect is cancelled. Conversely, when determining the SPE through the electric current inversion $\pm I_{e,y}$ method, it is summed up.
The offset is subtracted by using one measurement with magnetic field inversion method at saturation as a reference.
Also the spin Seebeck experiment is affected by a spurious voltage measured at Pt which is the reciprocal of the one found in the spin Peltier with current inversion. A very small transverse leaking heat flux may gives rise to electric effects in the nV range caused by the ordinary Seebeck effect due to the electric contacts between different metals. Again, this offset voltage is eliminated by using one magnetic field inversion point at saturation as a reference.

\section{Acknowledgement}
We thank Kelly Morrison at the University of Loughborough for the deposition of the Pt film. We thank Luca Martino and Federica Celegato (from Istituto Nazionale di Ricerca Metrologica) for technical support during the assembling of the measurement system.
One author (C.D.) thanks R. Meyer and B. Wenzel for technical support in sample preparation.

\section{Author information}
A.S., V.B., M.K. and M.P. devised the measurement technique and wrote the manuscript.
C.D. prepared the YIG sample, A.S. developed the measurement setup and characterized the device, V.B. developed the theoretical model.
All authors discussed the results, their physical interpretation and reviewed the manuscript.

\section{Additional information}
\subsection{Competing Interests:} The author(s) declare no competing interests.


\end{document}